\documentclass[prd,aps,preprint,tightenlines,superscriptaddress,nofootinbib,showpacs,showkeys,floatfix]{revtex4-2}
\usepackage{amsfonts}
\usepackage{array}
\usepackage{epsfig}
\usepackage{rotating}
\usepackage{multirow}
\usepackage{shuffle}
\usepackage{makeidx}
\usepackage{caption}
\usepackage{subcaption}
\usepackage{bigstrut}

\usepackage{amsmath}    
\usepackage{verbatim}   
\usepackage{color}      
\usepackage{colordvi}
\usepackage{hyperref}   

\usepackage{graphicx}

\usepackage{fancyvrb}

\definecolor{darkblue}{rgb}{0,0,0.5}
\definecolor{darkred}{rgb}{0.5,0,0}
\definecolor{darkgreen}{rgb}{0,0.5,0}

\newcommand{\fig}[1]{Figure~\ref{#1}}

\graphicspath{{../pics/}}
\DeclareGraphicsExtensions{.eps,.ps}

\usepackage{pslatex}
\usepackage{txfonts}
\usepackage[latin1]{inputenc}
\usepackage[T1]{fontenc}

\begin{document}


\begin{center}
  {\bf \Large
    Particle physics facing a pandemic
  }
  \vspace{1.25cm}

 {\large
   Adam~Kardos$^{\, a}$ 
   ,
   Sven-Olaf~Moch$^{\, b}$ 
   and
   Germ\'an~Rodrigo$^{\, c}$~\footnote{GR is currently on leave at the European Research Council Executive Agency, European
Commission, BE-1049 Brussels, Belgium. The views expressed are purely those of the writer and may not in any circumstances be regarded as stating an official position of the European Commission.}
   \\
 }
 \vspace{1.cm}
 {
   $^a$
   Department of Experimental Physics, Institute of Physics, Faculty of Science and Technology, 
   University of Debrecen\\
   4010 Debrecen, PO Box 105, Hungary\\[2ex]

   $^b$ II. Institut f\"ur Theoretische Physik, Universit\"at Hamburg \\
   Luruper Chaussee 149, D--22761 Hamburg, Germany \\[2ex]

   $^c$ Instituto de Fisica Corpuscular, Universitat de Valencia - Consejo Superior de Investigaciones Cient\'{\i}ficas, 
   Parc Cientific, E-46980 Paterna, Valencia, Spain
   \vspace{0.2cm}
 }
  \vspace{1.4cm}
\end{center}

\begin{center}
\textbf{Abstract}
\end{center}
  \textbf{Background:} Ordinary life as we knew it changed in March of 2020 due to the global COVID-19
  pandemic. While springtime in general was well awaited and regarded as a synonym for
  rejuvenation, March of 2020 on the other hand brought lock-down, curfews, home working and digital
  education to the lives of many. The particle physics community was not an exception:
  research institutes and universities introduced remote working and digital lecturing
  and all workshops, conferences and summer schools were cancelled, got postponed or 
  took place online.\\
 \textbf{Methods:} Using publicly available data from the INSPIRE and arXiv databases we investigate the effects of this
  dramatic change of life to the publishing trends of the high-energy physics
  community with an emphasis on particle phenomenology, theory and CERN's two major LHC experiments, ATLAS and CMS. 
  To get insights, we gather information about publishing trends in the last 20 years, ending by December 2021, and analyse it in detail.\\ 
  \textbf{Results:} Our analysis revealed that the publishing trend in particle physics was only been affected in a minor way.\\ 
  \textbf{Conclusions:} Publication data show that difficult times were successfully overcome and that the community even increased scientific output.

%

\newpage
\setcounter{footnote}{0}
\setcounter{page}{1}

\section*{Introduction}
\label{sec:intro}

In January 2020, the high-energy physics community was looking forward to another busy
year full of workshops, schools and conferences. The situation dramatically worsened in Europe by the end of February
and the final and lethal blow to all these live events happened in the middle
of March. 
The quick escalation of the situation was personally experienced by the
authors during the organization of the 
{\it PRecision Effective FIeld Theory School (PREFIT20)} 
from the 2\textsuperscript{nd} to the 13\textsuperscript{th} of March~\footnote{The {\it PREFIT20} school~\cite{prefit20} was organized as a joint event of the COST Actions {\it PARTICLEFACE}~\cite{particleface} and {\it VBSCan}~\cite{vbscan} at DESY in Hamburg.} 
and during a COST short-term scientific mission to Hamburg. 
The {\it PREFIT20} school had to be switched from lecturers being on-site 
to online presentations as events were running and turned out to be the last major live event in our field.
The short-term scientific mission looked like an ordinary trip 
during the out-bound journey (9\textsuperscript{th} of March), but one and a half weeks later (18\textsuperscript{th} of March) 
almost all flights were cancelled and the airport at Frankfurt became completely deserted, 
with all shops closed.
 
By the end of March, lock-down, working from home and online education became the standard. 
This also meant all conferences and workshops were cancelled or postponed until the end of summer. 
Within a couple of weeks the dominant platform to interact with colleagues became Zoom. 
The system got so wide-spread that term "Zoom meeting" became the 15\textsuperscript{th} most frequently used term in 2020~\cite{IrishTimes}.
In the particle physics community, collaborations spanning over several countries or continents are
quite common, having members from different countries in Europe or even from the Americas and Asia. Hence people were more-or-less familiar with virtual meeting
platforms and the technical transition had no major difficulties or problems. 
This was also witnessed
by the authors participating in several virtual collaboration meetings, seminars, online workshops and 
by giving lectures. 

Our experience suggests that
 although
distant collaborations had been set up and were working flawlessly well before the pandemic,
the situation changed because 
several institutions and universities in Europe introduced travelling

restrictions, and completely prohibited short visits and
brain-storming sessions during conference coffee breaks. 
Also, it is very common that new collaborations form spontaneously, a process
for which in-person meetings, workshops 
and structured programs at centers for scientific exchange like those offered by, e.g.,
Aspen, GGI, INT, KITP, MIAPP and MITP, play an important role~\footnote{
Aspen Center for Physics~\cite{aspen}, 
Galileo Galilei Institute (GGI) for Theoretical Physics~\cite{ggi}, 
Institute for Nuclear Theory (INT)~\cite{int}, 
Kavli Institute for Theoretical Physics (KITP)~\cite{kitp}, 
Munich Institute for Astro- and Particle Physics (MIAPP)~\cite{miapp}, 
Mainz Institute for Theoretical Physics (MITP)~\cite{mitp} provided.
}.
During a pandemic virtual presence and online platforms can keep an already working collaboration going,
but new collaborations become much harder to start due to the lack of personal contacts between potential
members. 
In the particle physics community well before the pandemic the usual habit was to approach people
personally during a workshop or conference and start an informal collaboration growing out of stimulating
discussions over coffee breaks.

The pandemic not only changed the professional lives of researchers but also their personal lives. This can also have consequences to scientific output and should not be disregarded.
With lockdown and online education introduced in schools researchers not only had to 
cope with new ways to keep their current collaborations going and to establish new ones 
despite the lack of all personal contacts but also had to prepare for lecturing using virtual classrooms. 
The latter aspect is of particular significance as the community generally regards chalk board lecturing as the best way to convey course material and scientific ideas 
\cite{Rudow,Waters644567}
. 
All of these factors have an effect on scientific research and it is not at all
clear how the community could adopt to new circumstances. 

In this paper we analyze the research output
in the form of papers and proceedings
of the high-energy particle physics phenomenology
community by using various open search engines to retrieve data on published papers during the pandemic and in the past
twenty years to get a better understanding of the true effects of the pandemic and to see what
kind of changes it induced in the community. To have a broader perspective we also collected data
related to the high-energy particle physics theory community and
the two major experiments, ATLAS and CMS, conducted at CERN's Large
Hadron Collider (LHC) as well.
We have selected these two experiments since these are the biggest ones in our community founded in
the middle of the 1990s, hence their publishing trends extends well before the pandemic.

\section*{Methods}

In order to extract publication data we used the search engines 
of two publicly available databases, which cover basically all relevant scientific publications 
(these include both peer-reviewed and non-peer-reviewed publications as well) 
in the field of high-energy physics: 
INSPIRE~\cite{inspire-hep} and arXiv~\cite{arxiv}. 
These engines allow for sophisticated searches by restricting hits through the use of many criteria, 
like paper categories
\footnote{The paper category represents a subfield in the high-energy physics community: hep-ph is destined for papers by the high-energy physics phenomenology community, hep-th for high-energy theoretical investigations and findings and hep-ex is exclusively used by the high-energy experiments publishing their results.}

(hep-ph, hep-th, hep-ex, etc.), 
by requiring papers to have a certain number of authors or 
by listing papers published or uploaded within a given period of time.
As these highly specialized tools have become quite mature it is possible to perform 
automatic queries by invoking the site's search engine directly with an API (Application Programming Interface).
By using a special URL, the search engine can be instructed to perform a search according to the
parameters specified through the URL and to return the result in some machine readable format.
For example, if publications that appeared in 2021 and categorized as hep-ph 
are extracted from INSPIRE the following URL should be invoked from the browser:

\begin{Verbatim}
https://inspirehep.net/api/literature?fields=hits.total&format=json&q=
  publication_info.year:2021%20AND%20arxiv_eprints.categories:hep-ph
\end{Verbatim}

The result is delivered in the \texttt{JSON} format, which is tolerably readable by the human eye but easily digestible
by the many commonly available interpreting languages, like \texttt{Python}. 
In our case only the total number of papers fulfilling certain criteria 
was needed, hence the method we have employed is:
\begin{enumerate}
  \item URL is constructed with all needed keywords.
  \item URL is invoked from \texttt{Python}.
  \item Result is delivered in \texttt{JSON} format.
  \item Total number of papers is extracted and written into data file.
  \item Histograms are created with \texttt{GNUplot}.
\end{enumerate}
In the case of the arXiv database we did not use the arXiv API. 
Instead, by selecting the advanced search option, 
the total number of records fulfilling the search criteria (for details see the corresponding section) were gathered manually.

In recent times a strong increase in conference proceedings can be seen in all the categories we have investigated.
In order to avoid this proliferation as much as possible and to get a clearer view of publishing trends 
we have filtered out all records without a DOI number. 
This method does not fully eliminate proceedings from the search results but restricts 
their number to a subset which is indeed published in one form or another. 
The filtering according to the DOI number was only possible with the INSPIRE API. 
Hence, we have validated our arXiv results by taking a look at the annual
statistics reported from both search engines to see if the presence of
proceedings in the arXiv search results affects the trends obtained with the INSPIRE API.

With the INSPIRE API we have collected annual statistics starting in the year 2001 and up to 2021. We have not considered 2022 because entries in INSPIRE do not acquire a DOI until they are published in a journal a few months later.
To get the annual number of papers we have requested that the field \texttt{publication\_info.year} includes the year
for which total number of publications are collected. 
Moreover, the document type had to be \texttt{article}, 
all the records were supposed to have a DOI number and to belong to the target category (hep-ph, hep-th or hep-ex)
for which we collected the data. 
In case of the hep-ex category we have only accepted papers by the ATLAS and
CMS collaborations and, hence, rejected any other. 
By varying the number of authors we were able to get statistics data for
various collaboration sizes.

In the case of the arXiv search engine our set of selection criteria was very modest. 
We have performed searches in the same three categories as in case of the
INSPIRE API, filtered records according to their submission year 
and, for a finer resolution, the month in which the submission happened.

\section*{Annual statistics using INSPIRE API}

For the long-term annual trend in publishing the INSPIRE API was used in the three different categories:
particle phenomenology (hep-ph), theory (hep-th) and experiment (hep-ex), 
the latter being restricted to the ATLAS and CMS collaborations only, and 
the results were filtered by requiring a valid DOI number attached to each record in order to minimize bias due to the proliferation of proceedings.
Specifically we are interested in the total number of papers on an annual basis and in the number of authors.

For the interpretation of the data collected in this way, it is also important to identify 
scientific events with potential impact on the publishing behavior of the high-energy physics community, 
such as new discoveries, the schedule of LHC operations, or other important developments.
For the categories we have analyzed events with potential impact are listed in Table 1. 
\begin{table}[!th]
  \centering
  \begin{tabular}{|c|c|}
    \hline\hline
    Event & Date \bigstrut\\
    \hline\hline
    LHC Run I & late 2009 - early 2013 \bigstrut\\
    \hline 
    OPERA announcement of measurement of superluminal neutrinos~\cite{OPERA:2011ijq} & 23 September 2011 \bigstrut\\
    \hline
    CERN announcement of Higgs boson discovery~\cite{Higgs2012} & 04 July 2012 \bigstrut\\
    \hline
    ATLAS and CMS presentation of diphoton excess at $750$ GeV~\cite{ATLAS-CONF-2015-081,CMS-PAS-EXO-15-004} & 15 December 2015 \bigstrut\\
    \hline
    LIGO announcement of first measurement of gravitational waves~\cite{LIGOScientific:2016aoc} & 11 February 2016 \bigstrut\\
    \hline
    LHC Run II & 2016 - 2018 \bigstrut\\
    \hline\hline
  \end{tabular}
  \caption{{\label{tbl:Events}} 
    Noteworthy events with potential impact on publishing trends and tendencies
    in the high-energy physics community since the start of LHC operations.}
\end{table}

\begin{figure}[!th]
  \centering
  \begin{subfigure}[!th]{0.6\textwidth}
    \includegraphics[width=1.0\textwidth]{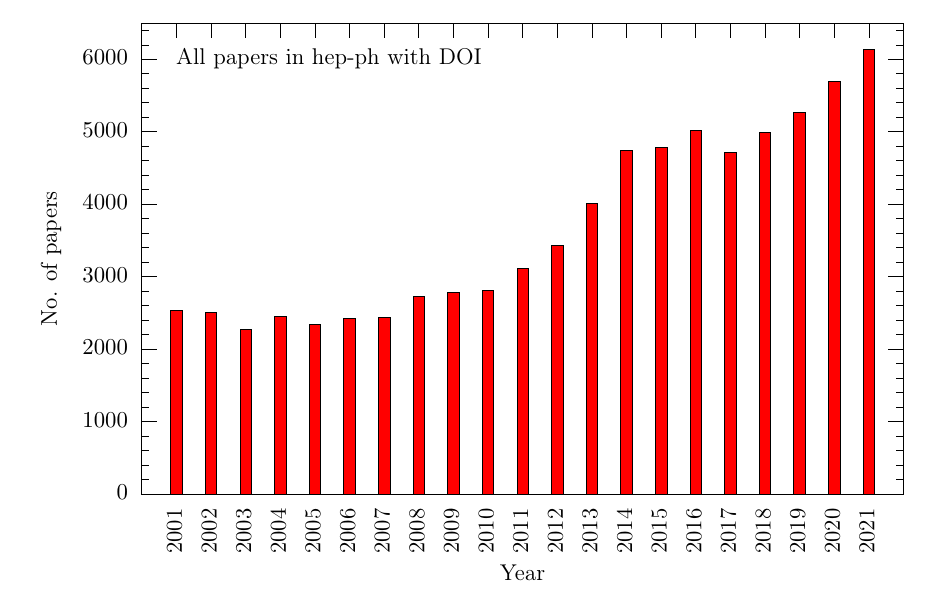}
    \caption{{\label{fig:inspire-annual-doi-ph}} Papers published with category label hep-ph.}
  \end{subfigure}
  \begin{subfigure}[!th]{0.6\textwidth}
    \includegraphics[width=1.0\textwidth]{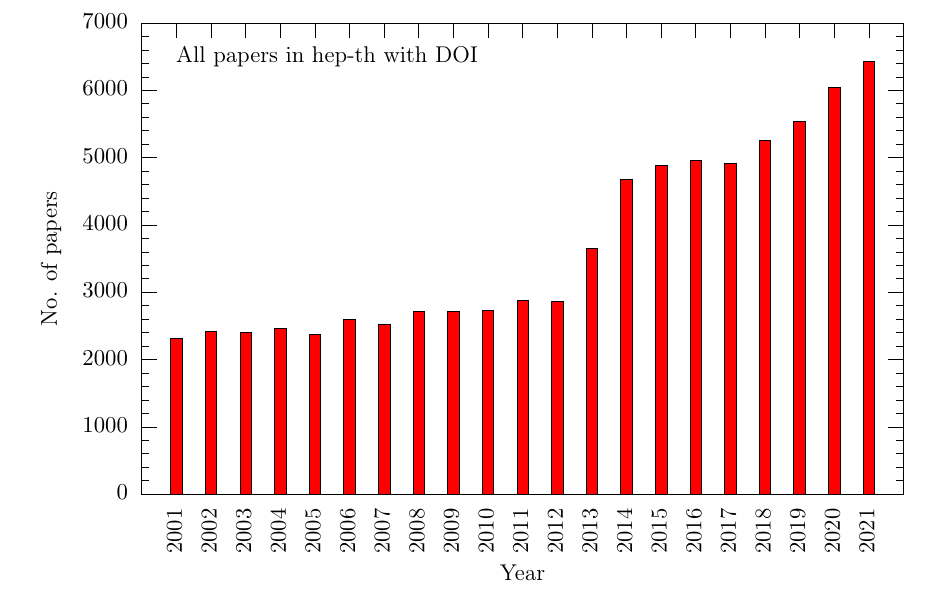}
    \caption{{\label{fig:inspire-annual-doi-th}} Papers published with category label hep-th.}
  \end{subfigure}
  \begin{subfigure}[!th]{0.6\textwidth}
    \includegraphics[width=1.0\textwidth]{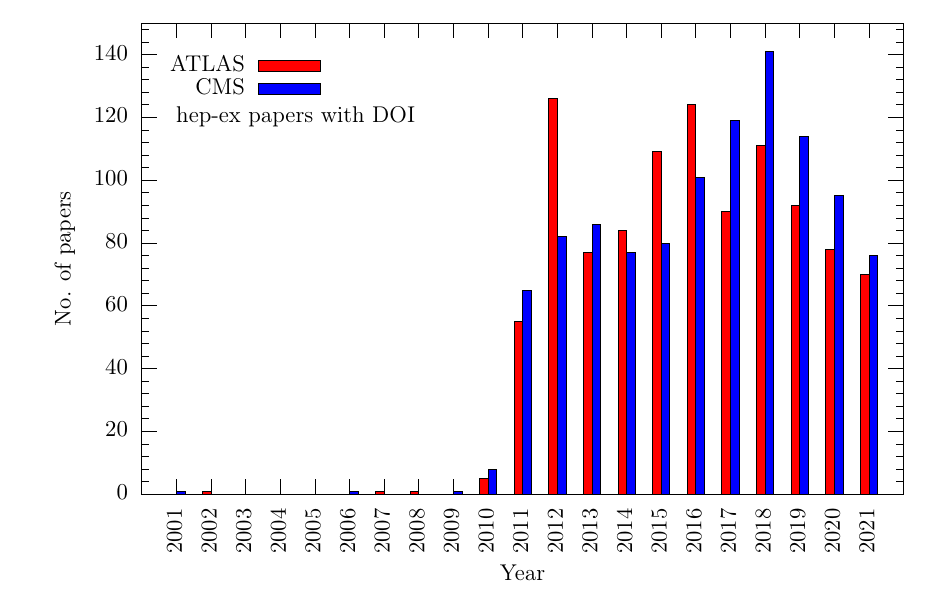}
    \caption{{\label{fig:inspire-annual-doi-ex}} Papers published with category label hep-ex by the ATLAS and CMS experiments.}
  \end{subfigure}
  \caption{{\label{fig:inspire-annual-doi}} Total number of published papers according to INSPIRE filtered
  using various arXiv category labels and requesting a valid DOI number.}
\end{figure}

\section*{Results}
\subsection*{Total number of papers}

The total number of published papers in the period 2001 to 2021 are depicted
in \fig{fig:inspire-annual-doi}. 
It is evident that the start of LHC
operations with Run I had a tremendous effect on publication activity in high-energy physics.
As it can be seen in \fig{fig:inspire-annual-doi-ph}, before 2010 the number of papers
produced in the phenomenology community remained at a steady level with a slight increase hardly discernable due to fluctuations. 
However, as first LHC results started to pour in the number of papers increased rapidly. 
A certain fraction of the increasing number of papers in 2011 can be also 
attributed to the OPERA announcement on the measurement of superluminal neutrinos~\cite{OPERA:2011ijq}.
In 2014 and with the end of Run I the increase flattens out, the annual
publication rate becomes roughly constant although with a different normalization,
roughly a factor of two compared to pre-LHC days.
A small peak can be seen in \fig{fig:inspire-annual-doi-ph} in 2016, 
which can be attributed to the $750$ GeV diphoton excess~\cite{ATLAS-CONF-2015-081,CMS-PAS-EXO-15-004}. 
The Run II of LHC has a less dramatic effect on the number of phenomenology papers and
it is also shifted by a year or so starting to ramp up only in 2018. 
The increase in the number of publications at the end of the LHC Run II did not finish in 2019, 
instead it is still visible in data taken for 2020 and 2021.
Knowing that the COVID-19 pandemic had its first real effects in March 2020 this is a remarkable observation.
The publishing trend did not change by the measures enforced on the phenomenology community but on the
contrary even with all difficulties induced by the pandemic the community flourished. 

A look at the high-energy theory category in \fig{fig:inspire-annual-doi-th} 
shows that the number of papers published grows only moderately until the year 2013, 
where a sudden increase happened in the normalization.
This change of normalization coincides with new achievements in multi-loop amplitude
calculations where the first publications appeared on arXiv in 2013 \cite{Henn:2013pwa,Henn:2013nsa} and completely changed
the way loop integrals are being calculated nowadays.
The real change in the trend appeared after the Run I starting in 2015 and 2016.
Then, a second, less dramatic increase starts in 2018, a trend in agreement with the previous LHC related ramp-up. 
The number of theory papers increases not during but just after the end of the LHC Run II. 
This could be a consequence of the lack of new physics found in both runs, 
leading the community to speculate about the lack of new signals. 

The annual number of published papers produced by the ATLAS and CMS
collaborations are shown in \fig{fig:inspire-annual-doi-ex} and it is evident 
that the publication activity really started around 2011 and 2012 once real data was taken and
subsequently analyzed.
As for Run I, ATLAS outnumbered CMS which started to catch up during Run II later on.
Also between the two runs the continuous analysis of recorded data produced a steady flow of published papers.
As more and more data has been analyzed the rate of papers decreases, 
which is already visible in the numbers recorded after 2019.

\begin{figure}[!th]
  \centering
  \begin{subfigure}[!th]{0.49\textwidth}
    \includegraphics[width=1.0\textwidth]{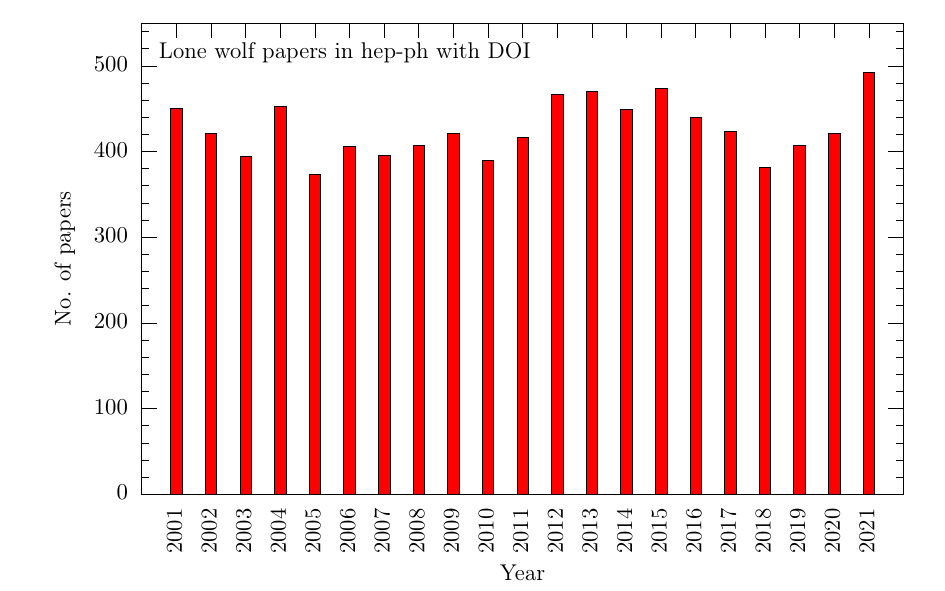}
    \caption{{\label{fig:inspire-annual-doi-ph-lonewolf}} Papers published
      by a single author with category label hep-ph.}
  \end{subfigure}
  \begin{subfigure}[!th]{0.49\textwidth}
    \includegraphics[width=1.0\textwidth]{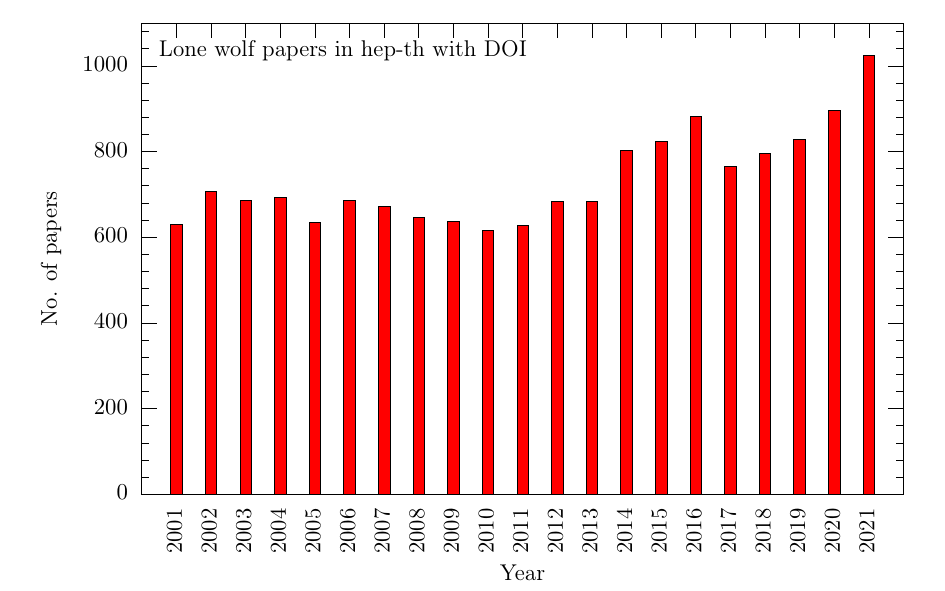}
    \caption{{\label{fig:inspire-annual-doi-th-lonewolf}} Papers published
      by a single author with category label hep-th.}
  \end{subfigure}
  \caption{{\label{fig:inspire-annual-doi-lonewolf}} Total number of published papers according to INSPIRE filtered
  using hep-ph (left) and hep-th (right) arXiv category labels, requesting a valid DOI number and with a single author.}
\end{figure}

\subsection{Number of authors}

In high-energy physics collaborations of many sizes exist. A pandemic can have different effects on different
collaboration sizes hence it is very informative to dissect the annual number of papers according to 
different collaboration size ranges. This partitioning can only affect phenomenology and theory papers
because in the hep-ex category we only consider ATLAS and CMS publications. 
The latter two collaborations have constant size since the start of LHC
operations in 2009 with approximately 3.000 authors in case of ATLAS and around
2.500 authors in case of CMS.

For the hep-ph and hep-th categories, the annual statistics for single-author papers is shown in \fig{fig:inspire-annual-doi-lonewolf}. 
Restricting ourselves to the phenomenology papers only, in \fig{fig:inspire-annual-doi-ph-lonewolf} 
it can be seen that the number of papers displays a rather steady trend 
with somewhat larger fluctuations due to the smaller amount of papers produced on an annual basis.
Due to higher fluctuations, possible trends are harder to identify but a slight increase can be observed 
at the start of LHC Run I in 2010 and ending in 2012. 
Recent years, starting with 2016, brought a steady decrease of number of papers with a slight increase starting 
in 2018 in accord with the end of LHC Run II.
This slight increase has continued until 2020 and has ramped up in 2021.

In \fig{fig:inspire-annual-doi-th-lonewolf},
single-author hep-th papers are listed as a function of publication year.
Before LHC Run I it follows a more-or-less constant trend with some fluctuations, even some decrease can
be seen ending in 2010. 
A ramp-up period can be identified starting in 2010 in line with the start of LHC operations resulting in a peak in 2016,
part of which can be attributed to the diphoton excess announcement in December
2015 and the LIGO announcement of first measurement of gravitational waves
in February 2016~\cite{LIGOScientific:2016aoc}.
Another period of increasing publication activity started in 2017 in line with
the end of the LHC Run II. 
The year 2020 witnessed an increasing publication activity that continued in 2021
for single-author papers in hep-th, so it seems that the pandemic was not a
significant factor in this respect.

\begin{figure}[!th]
  \centering
  \begin{subfigure}[!th]{0.49\textwidth}
    \includegraphics[width=1.0\textwidth]{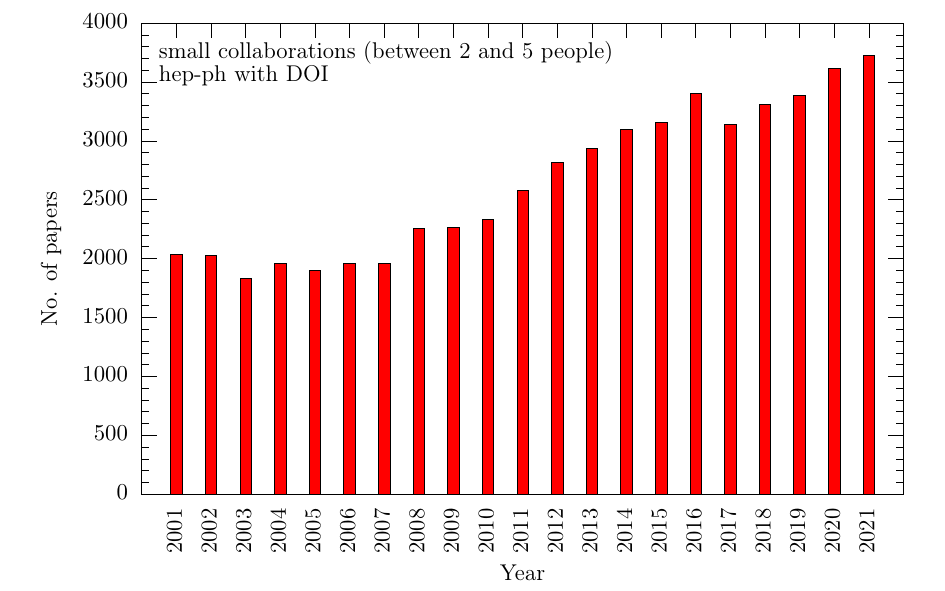}
    \caption{{\label{fig:inspire-annual-doi-ph-2-to-5}} Papers
      published by small-size collaborations 
      of two to five authors with category label hep-ph.}
  \end{subfigure}
  \begin{subfigure}[!th]{0.49\textwidth}
    \includegraphics[width=1.0\textwidth]{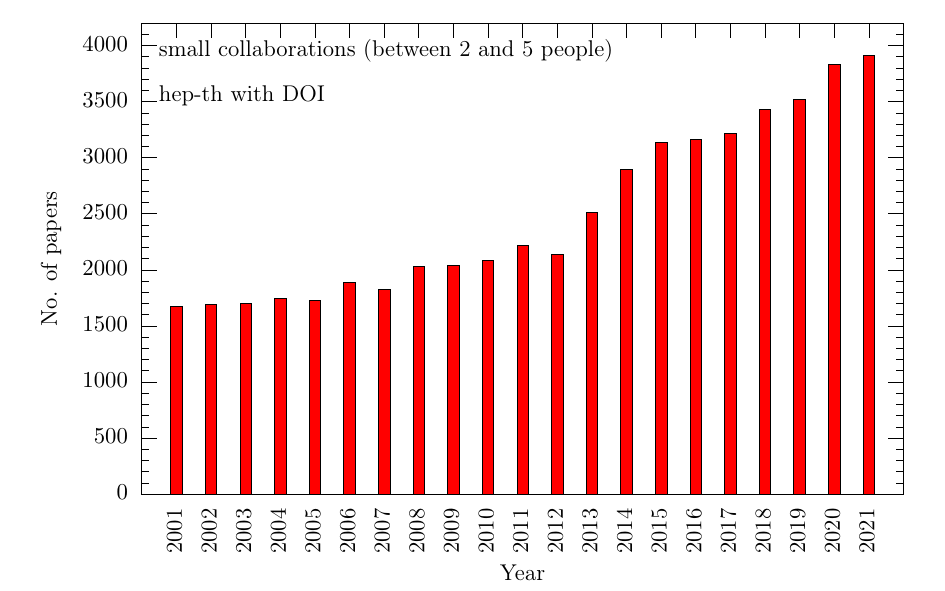}
    \caption{{\label{fig:inspire-annual-doi-th-2-to-5}} Papers
      published by small-size collaborations 
      of two to five authors with category label hep-th.}
  \end{subfigure}
  \caption{{\label{fig:inspire-annual-doi-2-to-5}} Total number of published papers according to INSPIRE filtered
  using hep-ph (left) and hep-th (right) arXiv category labels, requesting a valid DOI number and by collaborations with
  between two to five authors.}
\end{figure}

For small-size collaborations, which we define to consist of two to five authors, 
the number of published papers on an annual basis is plotted in \fig{fig:inspire-annual-doi-2-to-5} 
for both communities, phenomenology and theory. 
In \fig{fig:inspire-annual-doi-ph-2-to-5},
it is visible that before
the LHC Run I the phenomenology community more-or-less followed a steady publishing tendency with the same
amount of papers produced annually with some fluctuations. 
The start of Run I resulted in a steady increase in published
papers with a slight slow-down tendency as nearing the Run II. 
The year of 2016 witnessed a remarkably good year for publishing most probably
by the diphoton anomaly announced in late 2015 
\cite{ATLAS:2016gzy,CMS:2016xbb}
. 
Run II ignited one more increase which seems to continue.

On the other hand, in \fig{fig:inspire-annual-doi-th-2-to-5},  
publishing trends of small hep-th 
collaborations can be seen with a very similar behavior as in hep-ph, although
with a slightly elongated slow increase.
The big jump in publishing happens not through the Run I period but slightly after 
with no clear sign of superluminal neutrino or the diphoton excess announcements. 
The same trend is visible in recent years, as the numbers of publications grow
again with the end of Run II. 

\begin{figure}[!th]
  \centering
  \begin{subfigure}[!th]{0.49\textwidth}
    \includegraphics[width=1.0\textwidth]{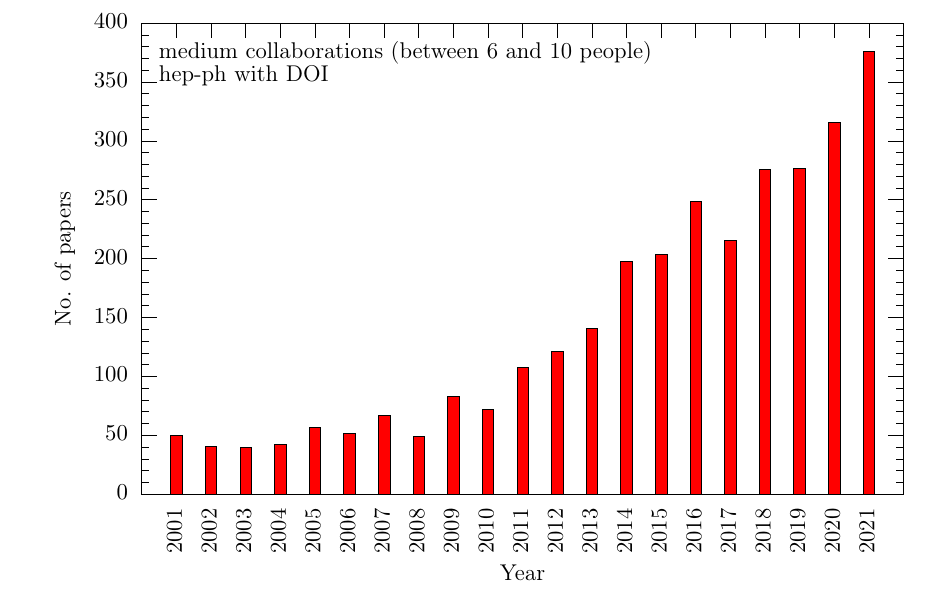}
    \caption{{\label{fig:inspire-annual-doi-ph-6-to-10}} Papers
      published by medium-size collaborations 
      of six to ten authors with category label hep-ph.}
  \end{subfigure}
  \begin{subfigure}[!th]{0.49\textwidth}
    \includegraphics[width=1.0\textwidth]{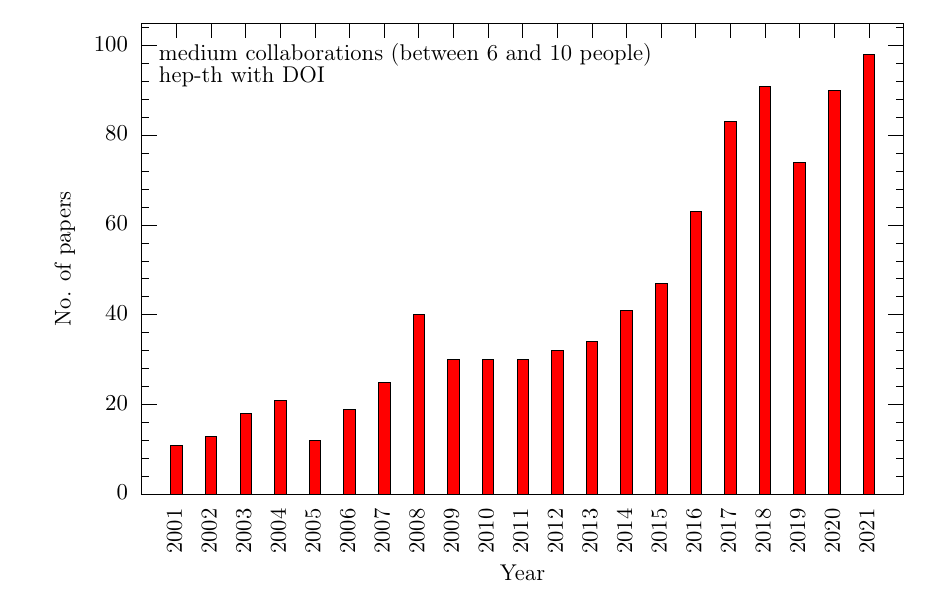}
    \caption{{\label{fig:inspire-annual-doi-th-6-to-10}} Papers
      published by medium-size collaborations 
      of six to ten authors with category label hep-th.}
  \end{subfigure}
  \caption{{\label{fig:inspire-annual-doi-6-to-10}} Total number of published papers according to INSPIRE filtered
  using hep-ph (left) and hep-th (right) arXiv category labels, requesting a valid DOI number and by collaborations with
  between six to ten authors.}
\end{figure}

Medium-size collaborations~\footnote{There is no solid definition what
  collaboration size can be considered medium-size. 
  In this paper drawing from our experience have settled for the number of
  authors in the range of six to ten in order to fit into this category.} 
are neither very common in phenomenology nor in theory as can be seen 
from the annual statistics in \fig{fig:inspire-annual-doi-6-to-10}.
It seems evident that the start of LHC operations during its Run I was a crucial factor in forging 
collaborations of this size since both in phenomenology and in theory the
number of papers published drastically increased throughout and after this
period. 
In the phenomenology statistics a clear peak is visible in 2016, 
correlated in time with the diphoton excess. 
In recent years, trends continue to be positive. 

\begin{figure}[!th]
  \centering
  \begin{subfigure}[!th]{0.49\textwidth}
    \includegraphics[width=1.0\textwidth]{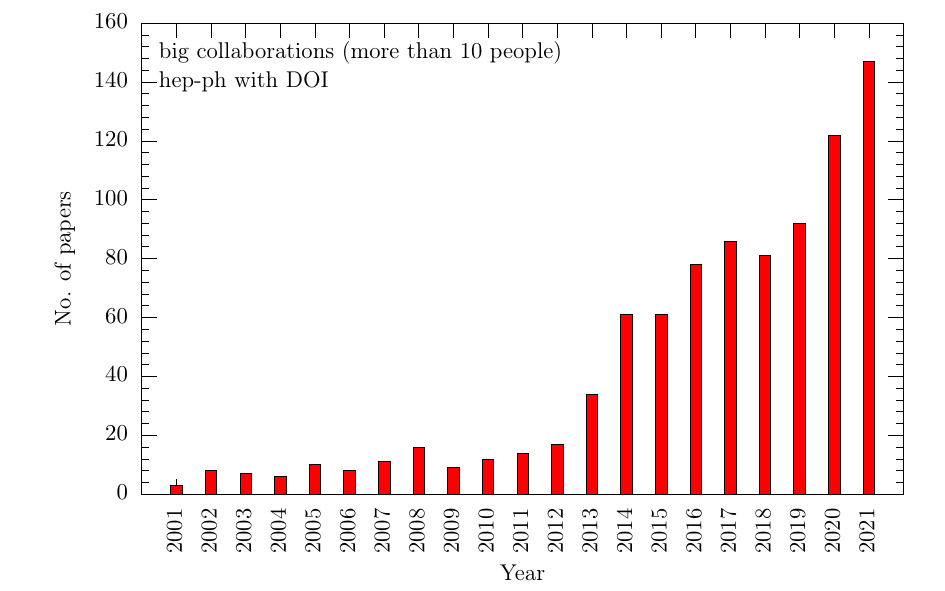}
    \caption{{\label{fig:inspire-annual-doi-ph-gt-10}} Papers published by large-size collaborations of more than ten authors with category label hep-ph.}
  \end{subfigure}
  \begin{subfigure}[!th]{0.49\textwidth}
    \includegraphics[width=1.0\textwidth]{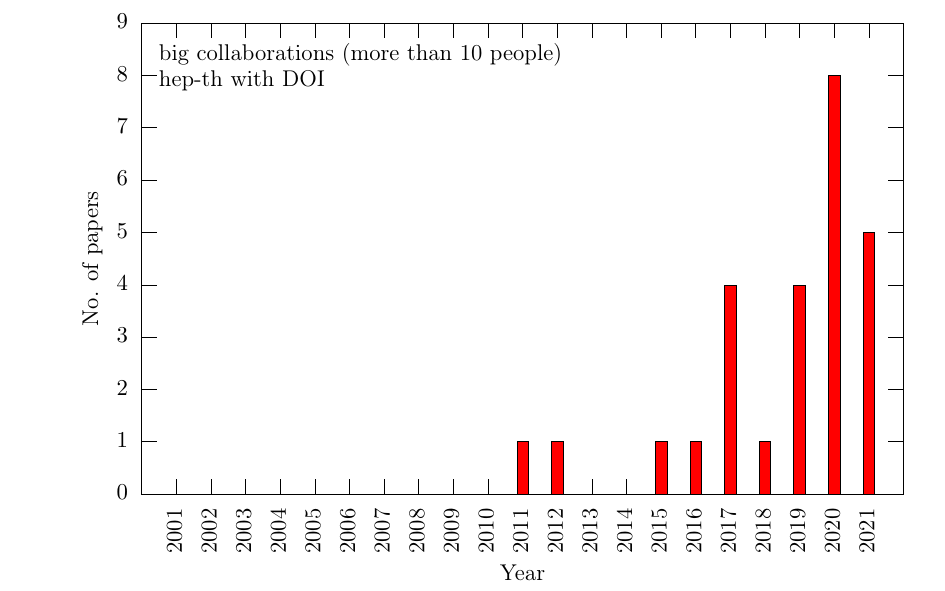}
    \caption{{\label{fig:inspire-annual-doi-th-gt-10}} Papers published by large-size collaborations of more than ten authors with category label hep-th.}
  \end{subfigure}
  \caption{{\label{fig:inspire-annual-doi-gt-10}} Total number of published papers according to INSPIRE filtered
  using hep-ph (left) and hep-th (right) arXiv category labels, requesting a valid DOI number by collaborations with
  more than ten authors.}
\end{figure}

Large-size collaborations~\footnote{As with medium-size collaborations there is no firm definition of what could be considered
  a large-size collaboration in these areas of academic research. Hence the
  lower bound for these collaborations of more than ten people reflects the authors' judgment 
  of what can be regarded as big in our communities.} 
of more than ten authors involved do not have a long history in the analyzed two
communities as can be seen in \fig{fig:inspire-annual-doi-gt-10}.
In the theory community even nowadays these can be regarded as scarce 
with less than a dozen papers produced annually. 
As \fig{fig:inspire-annual-doi-ph-gt-10}
shows the advent of LHC physics and the first results from LHC Run I catalyzed the formation of these kind of collaborations.
In comparison, less than twenty such papers were published on an annual basis before the LHC era. 
The past recent years however do witness a rising trend of publications from 
large-size collaborations in the hep-ph category, culminating in an enormous
peak in 2020 (more than 20\% increase compared to 2019) and 2021. 
This dramatic increase in the number of publications by large-size collaborations in
phenomenology can be explained by the extraordinary performance of LHC during and since Run I. 
The high-quality data created a strong driving force for precise predictions,
which, on the other hand, require a significant amount of collaborators.
Consequently, the demand for state-of-the-art predictions can often only be
fulfilled within a reasonable time-frame by joining efforts, 
hence large(r) collaborations have been and are being formed. 
The peaking behavior can also be seen in the publication trend 
in the hep-th community but since the number of publications there is smaller
by a factor of two the increase in 2020 compared to 2019, and the decrease in 2021, can be partially
attributed to fluctuations. 
However, it is clear from the trend (at least for the hep-ph community) that the pandemic did not have any effect on the publishing
trend. On the contrary, the past two years were fruitful for producing papers for this size of collaborations.

\bigskip

Judging from these annual statistics, it seems that the effect of the pandemic
on the three communities in high-energy particle physics (hep-ph, hep-th and
ATLAS, CMS in hep-ex) is at most moderate and in the worst case resulted in a stagnation in the
publishing trends or a very slight decrease. 
Collaborations of larger sizes seem to be immune to the COVID-19 restrictions, 
being able to publish with the same or even higher frequency. 
In the case of the ATLAS and CMS collaborations the 
overall trend for the recent years shows a decreasing rate of publications and the data for 2020 and 2021 also mirror this well-established trend, which is likely to change after the start of the LHC Run III in 2022..

\section*{Monthly statistics using arXiv}

\begin{figure}[!th]
  \centering
    \includegraphics[width=0.6\textwidth]{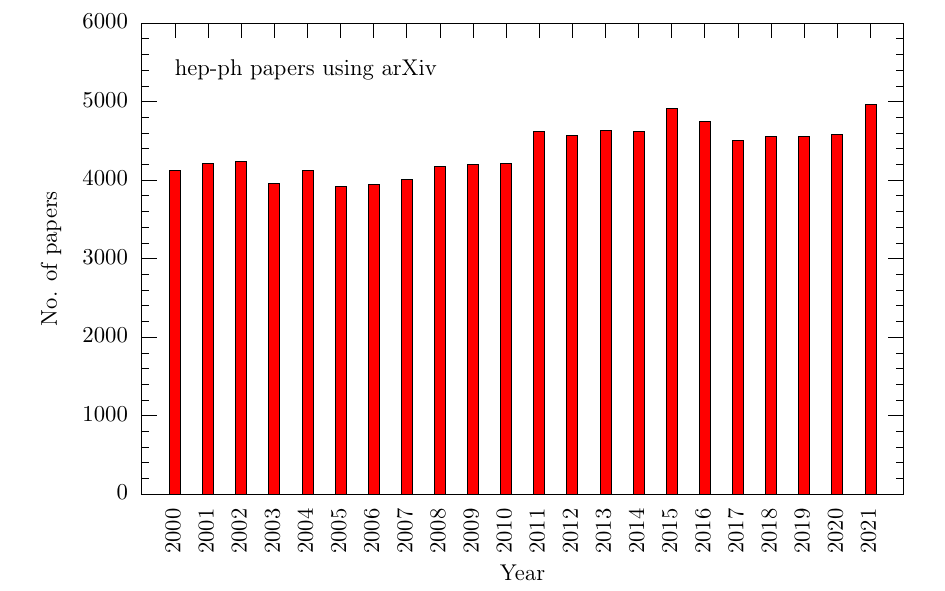}
  \caption{{\label{fig:arxiv-annual-ph}} Annual submission statistics in the hep-ph category using
  the arXiv search engine.}
\end{figure}

The search engine of arXiv allows us to extract publishing data not only on an annual
but also on a monthly basis, i.e., we can get some insights on the monthly distribution of publications. 
The engine does not just restrict the search to published papers, but even with conference proceedings contained in the collected samples, it is interesting to look at publication and arXiv submission trends at a finer resolution.
In order to convince ourselves that statistics gathered on a monthly basis including conference proceedings still
can be used to identify trends in publishing,
the total annual statistics collected via arXiv 
can be compared with the histograms in the previous section based on the data of the INSPIRE API. 
As an example of this, cross-check annual submission statistics were collected in the 
hep-ph category on arXiv and the resulting histogram is plotted in 
\fig{fig:arxiv-annual-ph}. 
Comparing this figure to \fig{fig:inspire-annual-doi-ph} 
shows that the trends are the same, although the overall normalization is different. 
Before the LHC Run I period both statistics follow a more-or-less constant trend 
with an increase in the number of publications due to the LHC Run I. 
Even the publishing peak in 2016 is visible with arXiv results and the increasing frequency of recent years
can be identified. These findings give us enough confidence to take a look at the monthly charts 
obtained from arXiv and to search for tendencies in them.

\begin{figure}[!th]
  \centering
  \begin{subfigure}[!th]{0.7\textwidth}
    \includegraphics[width=1.1\textwidth]{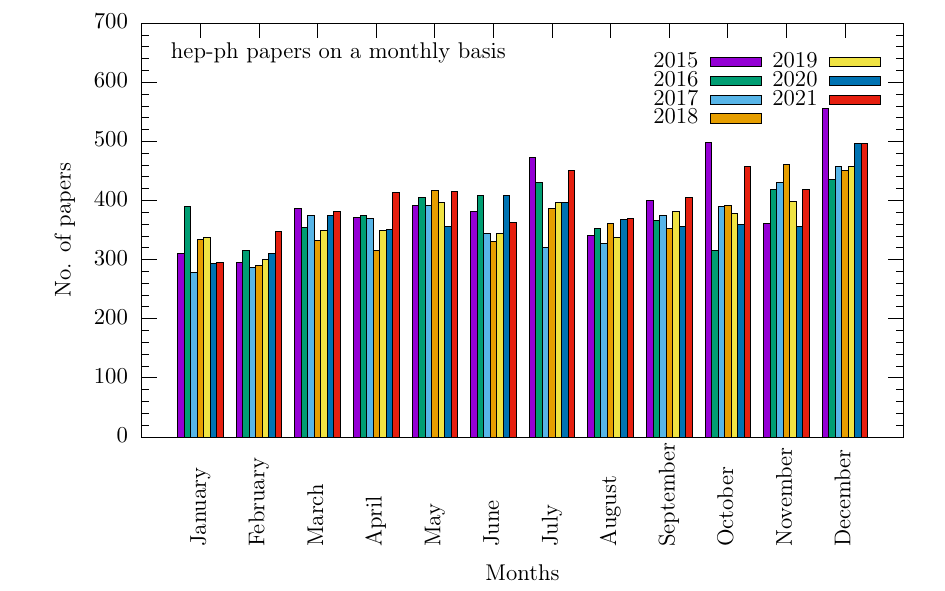}
    \caption{{\label{fig:arxiv-monthly-ph}} Submissions to arXiv on a monthly basis in the category label hep-ph.}
  \end{subfigure}
  \begin{subfigure}[!th]{0.7\textwidth}
    \includegraphics[width=1.1\textwidth]{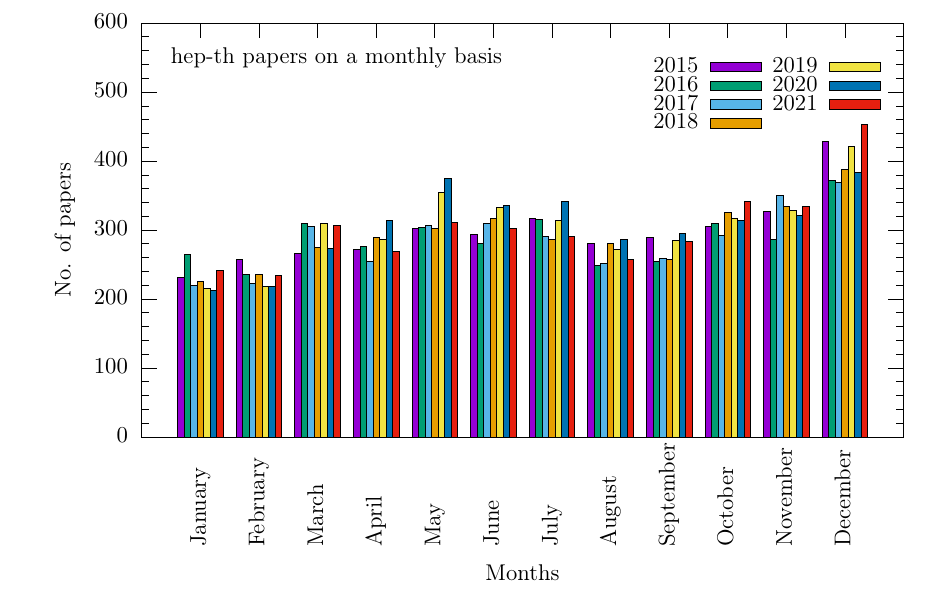}
    \caption{{\label{fig:arxiv-monthly-th}} Submissions to arXiv on a monthly basis in the category label hep-th.}
  \end{subfigure}
  \caption{{\label{fig:arxiv-monthly}} Submission to arXiv in the hep-ph (up) and hep-th (down) 
  categories between 2015 and 2021. For further details see the main text.}
\end{figure}

In \fig{fig:arxiv-monthly}
we have plotted the monthly statistics for seven consecutive years, from 2015 to 2021, for both hep-ph
and hep-th. As it can be seen from both plots the publishing trend is relatively constant for
phenomenology papers but for theory papers a clear pattern emerges. 
The publishing frequency is systematically lower in the first two months of the year 
and increases to a local maximum in the summer but just to turn back and reach a second minimum 
in August and September and then it ramps up to reach the maximum frequency as the year ends. 
The histograms contain both 2020 and 2021 results and these data fit perfectly well 
into the general trend of recent years. 
In order to create these histograms we have used the original submission date
for the records to filter results. 
In this way it is easy to see for example that the $750$ GeV diphoton excess
really created a swell of publications~\footnote{In the community this effect is sometimes
  referred to as ``chasing the ambulance''.}, 
both in phenomenology and theory papers, resulting in an excess of 15-20 \% of publishing in December 2015
with a significant  surplus carried over to January 2016 (more than 30\% for phenomenology and 17\% for theory compared to the
same period of previous year).

From these statistics it seems that the pandemic did little to change the already settled publishing
trends and at times it helped overshoot the same month from previous years. We cannot identify any dramatic
increase nor dramatic decrease in the publication rates.

\section*{Conclusions}

Despite the pandemic, the high-energy physics community continued with
the increasing trend in the number of publications that started with 
the operation of the LHC.
At the same time the face-to-face interaction remains a vital component in the process of
exchanging scientific ideas, for work in existing collaborations and for the start of new ones.
Here, workshops, conferences and topical programs at centers for scientific exchange play an important role 
by creating suitable environments and atmospheres.
The complete lack of such interactions could have a negative impact.  
While the various virtual presence platforms seem suitable and efficient for administrative
purposes and supersede in-person meetings, they cannot serve as a substitute
for the true melting pots of new ideas during social events.
Travel and workshop attendance can also help to focus on research in the
absence of the usual other errands, lectures and administrative burdens.

In summary, the particle physics community successfully adapted to the COVID-19 pandemic and the
associated restrictions for the time being, but to avoid
complete stagnation and to really progress towards a bright future 
the high-energy physics community needs to restart in-person events, scientific visits and black board discussions.

\section*{Data availability}

Repository: Hepstat: Particle physics facing a pandemic - dataset
\href{https://zenodo.org/record/7392919}{https://doi.org/10.5281/zenodo.7392919}
This project contains the following underlying data:
\begin{itemize}
  \item \texttt{arxiv} (containing annual and monthly statistics extracted from arXiv)
  \item \texttt{gnuplot} (contains all the \texttt{GNUplot} scripts used to create all the figures of the paper in two subfolders: one for arXiv and one for the INSPIRE statistics)
  \item \texttt{inspire} (contains the annual statistics obtained from INSPIRE)
  \item \texttt{.py} files:
  \begin{itemize}
    \item \texttt{arxiv-annual.py}: this script is used to extract annual statistics from arXiv
    \item \texttt{arxiv.py}: this script is used to obtain the monthly statistics from arXiv
    \item \texttt{inspire.py}: this script is used to get the various annual statistics from INSPIRE
  \end{itemize}
\end{itemize}

All data has been extracted from open sources and underlying results and all computer code are available on \texttt{GitHub} \cite{hepstat}

Data are available under the terms of a \href{https://creativecommons.org/publicdomain/zero/1.0/}{CC0} licence and no additional source data are required.

\section*{Acknowledgements}

This work is supported by the COST Action CA16201 PARTICLEFACE, by the Spanish Government (MCIN/AEI/10.13039/501100011033)  
Grant No. PID2020-114473GB-I00 and Generalitat Valenciana  (Grant No. PROMETEO/2021/071),
by grant K 125105 of the National Research, Development and Innovation Fund in Hungary and by 
the \'UNKP-21-Bolyai+ New National Excellence Program of the Ministry 
for Innovation and Technology from the source of the National Research, Development and Innovation Fund.
AK kindly acknowledges financial support from the Bolyai Fellowship programme of the Hungarian
Academy of Sciences.

\section*{Ethics and consent}

No ethical approval or consent was required.

\bibliographystyle{JHEP}

\providecommand{\href}[2]{#2}\begingroup\raggedright\begin{thebibliography}{10}

\bibitem{prefit20}
{\it PREFIT20: PRecision Effective FIeld Theory School},
  \url{https://indico.cern.ch/event/817757}.

\bibitem{particleface}
{COST Action CA16201}, \emph{{\it PARTICLEFACE - Unraveling new physics at the
  LHC through the precision frontier}},  \url{https://particleface.eu}.

\bibitem{vbscan}
{COST Action CA16108}, \emph{{\it VBSCan - Vector Boson Scattering Coordination
  and Action Network}},  \url{https://vbscanaction.web.cern.ch}.

\bibitem{IrishTimes}
{Irish Times}, \emph{{\it What's~the~top~word~of~2020?
  Oh,~you~guessed~that~one~easily}},
  \url{https://www.irishtimes.com/culture/what-s-the-top-word-of-2020-oh-you-guessed-that-one-easily-1.4298315}.

\bibitem{aspen}
{\it The Aspen Center for Physics}, \url{https://www.aspenphys.org/}.

\bibitem{ggi}
{\it The Galileo Galilei Institute (GGI) for Theoretical Physics},
  \url{https://www.ggi.infn.it/}.

\bibitem{int}
{\it The Institute for Nuclear Theory (INT)},
  \url{http://www.int.washington.edu/}.

\bibitem{kitp}
{\it The Kavli Institute for Theoretical Physics (KITP)},
  \url{https://www.kitp.ucsb.edu/}.

\bibitem{miapp}
{\it The Munich Institute for Astro- and Particle Physics (MIAPP)},
  \url{https://www.munich-iapp.de/}.

\bibitem{mitp}
{\it The Mainz Institute for Theoretical Physics (MITP)},
  \url{https://www.mitp.uni-mainz.de/}.

\bibitem{Rudow}
F.J.E.~Rudow,~S.~R., \emph{Pointing with power or creating with chalk},
  \href{https://dx.doi.org/10.19030/cier.v8i3.9344}{\emph{Contemporary Issues
  in Education Research (CIER)} (2015)}
  [\href{https://arxiv.org/abs/https://www.biorxiv.org/content/early/2019/05/27/644567.full.pdf}{{\tt
  https://www.biorxiv.org/content/early/2019/05/27/644567.full.pdf}}].

\bibitem{Waters644567}
C.M.~Waters, \emph{Rock the chalk: A five-year comparative analysis of a large
  microbiology lecture course reveals improved outcomes of chalk-talk compared
  to powerpoint}, \href{https://dx.doi.org/10.1101/644567}{\emph{bioRxiv}
  (2019)}
  [\href{https://arxiv.org/abs/https://www.biorxiv.org/content/early/2019/05/27/644567.full.pdf}{{\tt
  https://www.biorxiv.org/content/early/2019/05/27/644567.full.pdf}}].

\bibitem{inspire-hep}
{INSPIRE Collaboration}, \emph{{INSPIRE}},  \url{https://inspirehep.net/}.

\bibitem{arxiv}
{Cornell University}, \emph{{arXiv}},  \url{https://arxiv.org/}.

\bibitem{OPERA:2011ijq}
{\scshape OPERA} collaboration, T.~Adam et~al., \emph{{Measurement of the
  neutrino velocity with the OPERA detector in the CNGS beam}},
  \href{https://dx.doi.org/10.1007/JHEP10(2012)093}{\emph{JHEP} {\bf 10} (2012)
  093} [\href{https://arxiv.org/abs/1109.4897}{{\tt arXiv:1109.4897}}].

\bibitem{Higgs2012}
{\scshape ATLAS and CMS} collaboration, F.~Gianotti and J.~Incandela,
  \emph{{Higgs announcement seminar on 4 July 2012}},
  \url{https://videos.cern.ch/record/1459513}, Jul, 2012.

\bibitem{ATLAS-CONF-2015-081}
{\scshape ATLAS} collaboration, \emph{{\it {Search for resonances decaying to
  photon pairs in 3.2 fb$^{-1}$ of $pp$ collisions at $\sqrt{s}$ = 13 TeV with
  the ATLAS detector}}},  preprint {\tt ATLAS-CONF-2015-081}, 2015,
  \url{http://cds.cern.ch/record/2114853}.

\bibitem{CMS-PAS-EXO-15-004}
{\scshape CMS} collaboration, \emph{{Search for new physics in high mass
  diphoton events in proton-proton collisions at $\sqrt{s} = 13$ TeV}},
  preprint {\tt CMS-PAS-EXO-15-004}, 2015,
  \url{https://cds.cern.ch/record/2114808}.

\bibitem{LIGOScientific:2016aoc}
{\scshape LIGO Scientific, Virgo} collaboration, B.P.~Abbott et~al.,
  \emph{{Observation of Gravitational Waves from a Binary Black Hole Merger}},
  \href{https://dx.doi.org/10.1103/PhysRevLett.116.061102}{\emph{Phys. Rev.
  Lett.} {\bf 116} (2016) 061102} [\href{https://arxiv.org/abs/1602.03837}{{\tt
  arXiv:1602.03837}}].

\bibitem{Henn:2013pwa}
J.M.~Henn, \emph{{Multiloop integrals in dimensional regularization made
  simple}},
  \href{https://dx.doi.org/10.1103/PhysRevLett.110.251601}{\emph{Phys. Rev.
  Lett.} {\bf 110} (2013) 251601} [\href{https://arxiv.org/abs/1304.1806}{{\tt
  arXiv:1304.1806}}].

\bibitem{Henn:2013nsa}
J.M.~Henn, A.V.~Smirnov and V.A.~Smirnov, \emph{{Evaluating single-scale and/or
  non-planar diagrams by differential equations}},
  \href{https://dx.doi.org/10.1007/JHEP03(2014)088}{\emph{JHEP} {\bf 03} (2014)
  088} [\href{https://arxiv.org/abs/1312.2588}{{\tt arXiv:1312.2588}}].

\bibitem{ATLAS:2016gzy}
{\scshape ATLAS} collaboration, M.~Aaboud et~al., \emph{{Search for resonances
  in diphoton events at $\sqrt{s}$=13 TeV with the ATLAS detector}},
  \href{https://dx.doi.org/10.1007/JHEP09(2016)001}{\emph{JHEP} {\bf 09} (2016)
  001} [\href{https://arxiv.org/abs/1606.03833}{{\tt arXiv:1606.03833}}].

\bibitem{CMS:2016xbb}
{\scshape CMS} collaboration, V.~Khachatryan et~al., \emph{{Search for Resonant
  Production of High-Mass Photon Pairs in Proton-Proton Collisions at $\sqrt s$
  =8 and 13 TeV}},
  \href{https://dx.doi.org/10.1103/PhysRevLett.117.051802}{\emph{Phys. Rev.
  Lett.} {\bf 117} (2016) 051802} [\href{https://arxiv.org/abs/1606.04093}{{\tt
  arXiv:1606.04093}}].

\bibitem{hepstat}
{A.~Kardos, S.O.~Moch and G.~Rodrigo}, \emph{Hepstat: Particle physics facing a pandemic - dataset},
  \url{https://doi.org/10.5281/zenodo.7392919}.

\end{thebibliography}\endgroup
\providecommand{\href}[2]{#2}\begingroup\raggedright\endgroup

\end{document}